\documentclass[a4paper]{appolb}

\usepackage{dcolumn}
\usepackage{wrapfig}
\usepackage{epsfig}

\def\Journal#1#2#3#4{{#1} {\bf #2} (#3) #4}

\def\PLB{{\em Phys. Lett.}   {\bf B}}

\def\PRD{{\em Phys. Rev.}    {\bf D}}
\def\ZPC{{\em Z. Phys.}      {\bf C}}
\def\EJC{{\em Eur. Phys. J.} {\bf C}}

\newcommand{\pom}{{I\!\!P}}

\newcommand{\dd}{\mathrm{d}}

\begin{document}

\title{
NLO QCD Fit to H1 Diffractive DIS Data
\thanks{Talk given at 10th Intl. Workshop on Deep Inelastic Scattering 
(DIS 2002), Cracow }
}

\author{
Frank-Peter Schilling (for the H1 Collaboration)
\address{DESY (Hamburg, Germany)}
}

\maketitle

\begin{abstract}
  A new NLO DGLAP QCD fit to recent inclusive diffractive DIS data
  from the H1 collaboration is presented.  Diffractive parton
  distributions are extracted, including their experimental and
  theoretical uncertainties.  The parton distributions are used for
  comparisons with recent diffractive final state data from HERA and
  the TEVATRON.
\end{abstract}

%\PACS{(not yet)}
  
\section{Introduction}

Measurements of the cross section $\sigma^D(x_\pom,\beta,Q^2)$ for
diffractive deep inelastic scattering (DIS) processes of the type
$ep\rightarrow eXp$ by H1 at HERA have reached high precision
\cite{laycock}.  Here, $Q^2$ is the photon virtuality, $x_\pom$ is the
longitudinal momentum fraction of the diffractive exchange with
respect to the proton and $\beta=x/x_\pom$ is the exchange's
longitudinal momentum fraction of the quark which couples to the
virtual photon.  The recent H1 data are presented in terms of a
reduced cross section $\sigma_r^D$, integrated over $t$, the
4-momentum transfer squared at the proton vertex,
\begin{equation}
\frac{d^3\sigma^D}{\dd x_\pom \ \dd \beta \ \dd Q^2} = \frac{4\pi \alpha^2}{\beta Q^4}\left ( 1 - y + \frac{y^2}{2} \right ) \sigma_{r}^{D(3)}(x_\pom,\beta,Q^2) \ ,
\end{equation}
($y=Q^2/xs$ is the inelasticity) which is expressed in terms of the
diffractive structure functions $F_2^D$ and the longitudinal $F_L^D$
as
\begin{equation}
\sigma_r^{D(3)} = F_2^{D(3)} - \frac{y^2}{1+(1-y)^2} F_L^{D(3)} \ .
\end{equation}
The proof \cite{collins} of QCD hard scattering factorization in
diffractive DIS justifies to express $\sigma_r^{D(3)}$, at fixed
$x_\pom$, as a convolution of diffractive parton distributions
(dpdf's) $f_i^D$ and partonic cross sections $\hat{\sigma}^{\gamma^* i}$:
\begin{equation}
\frac{\dd \sigma_r^{D(3)}(x,Q^2,x_\pom)}
{\dd x_\pom } \ = \
\sum_i \int_x^{x_\pom}\dd \xi \
\hat{\sigma}^{\gamma^*i}(x,Q^2,\xi) \
f_i^D(\xi,Q^2,x_\pom) \ .
\label{equ:diffpdf}
\end{equation}
The $f_i^D$ should obey the DGLAP evolution equations and the
$\hat{\sigma}^{\gamma^* i}$ are the same as for standard DIS.  In a
next-to-leading order (NLO) DGLAP QCD fit, dpdf's are determined from
the recent $\sigma_r^{D(3)}$ data from H1. Factorization in
diffraction is tested by using these dpdf's for comparisons with
diffractive final state data from HERA and the TEVATRON.

\section{The NLO QCD fit}

In the fit to the H1 $\sigma_r^{D(3)}$ data, the shapes of the dpdf's
are assumed to be independent of $x_\pom$ (``Regge factorization''),
which is supported by the data. The $x_\pom$ dependence is
parameterized as
\begin{equation}
f_\pom(x_\pom) = \int \dd t \ x_\pom^{1-2\alpha_\pom(t)}e^{B_\pom}t \ ; \qquad
\alpha_\pom(t)=\alpha_\pom(0)+\alpha'_\pom t \ ,
\end{equation}
where $\alpha_\pom(0)=1.173\pm 0.018$ as obtained from a fit to the
$x_\pom$ dependence of the data\footnote{For the values of
  $\alpha'_\pom$ and $B_\pom$ and the assumed uncertainties see
  \cite{f2deps}.}. A sub-leading exchange contribution (parameterized
using a pion pdf) is included but 
found to be negligibly small for
$x_\pom<0.01$.  The diffractive exchange is parameterized by a
light flavour singlet\footnote{$u=d=s=\bar{u}=\bar{d}=\bar{s}$ is
  assumed.}  and a gluon distribution at a starting scale $Q_0^2=3
\rm\ GeV^2$ using Chebychev polynomials.  No momentum sum rule is
imposed.  Heavy quarks are treated in the massive scheme with
$m_c=1.5\pm0.1 \rm\ GeV$. The strong coupling is set via $\Lambda^{\rm
  \overline{MS}}_{\rm QCD}=200\pm30 \rm\ MeV$.

The NLO DGLAP evolution equations are used. The QCD fit program is the
one used in \cite{f2}, extended for diffraction. In the fit, the
experimental systematic errors of the data points and their
correlations are propagated to obtain error bands for the resulting
dpdf's. In addition, a theory error is estimated by variations of
$\Lambda_{\rm QCD}$, $m_c$ and the parameterization of the $x_\pom$
dependences.

The data used in the fit cover $x_\pom<0.05$, $0.01\leq\beta\leq0.9$
and $M_X>2 \rm\ GeV$, where the latter cut on the diffractive mass
$M_X$ is applied to justify a leading twist approach. The NLO fit is
performed to $\sigma_r^D$.
However, due to the $y$ range of the data there is presently no direct
sensitivity to $F_L^D$.  Two data sets are used in the fit: (1) the
recent H1 preliminary data \cite{f2deps} in the range $6.5\leq Q^2\leq
120 \rm\ GeV^2$ (284 points) and (2) H1 preliminary data at higher
$Q^2$ \cite{f2dhiq2} ($200 \leq Q^2 \leq 800 \rm\ GeV^2$, 29 points).
The $\chi^2$ for the central NLO fit is $308.7$ for $306$ degrees of
freedom and 7 free parameters (3+3 parameters for the singlet and
gluon distributions, 1 normalization for the sub-leading exchange
contribution).

\begin{figure}[ht]
\centering
\epsfig{bbllx=0,bburx=510,bblly=50,bbury=550,
file=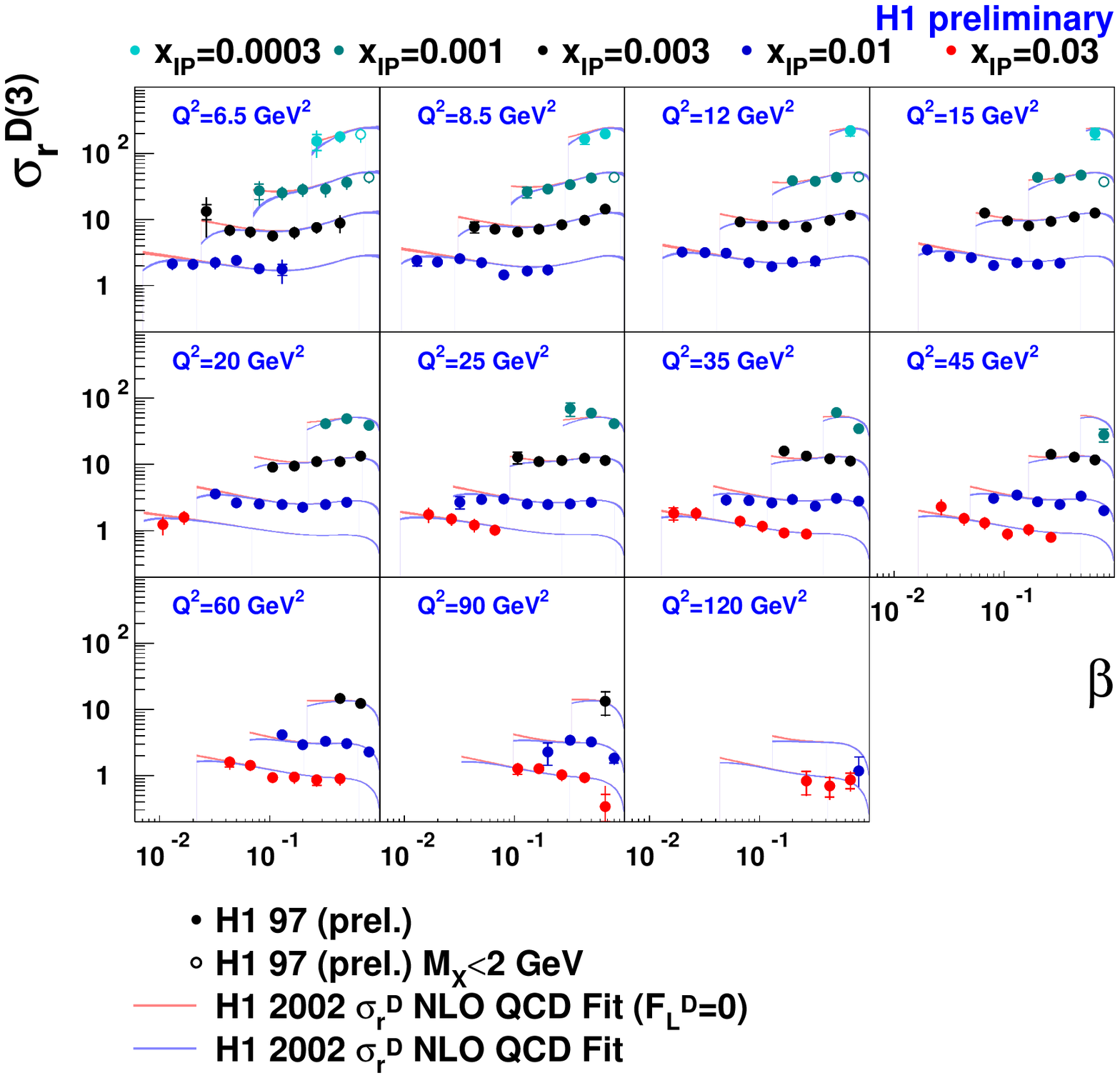,width=0.49\linewidth}
\epsfig{bbllx=0,bburx=510,bblly=50,bbury=550,
file=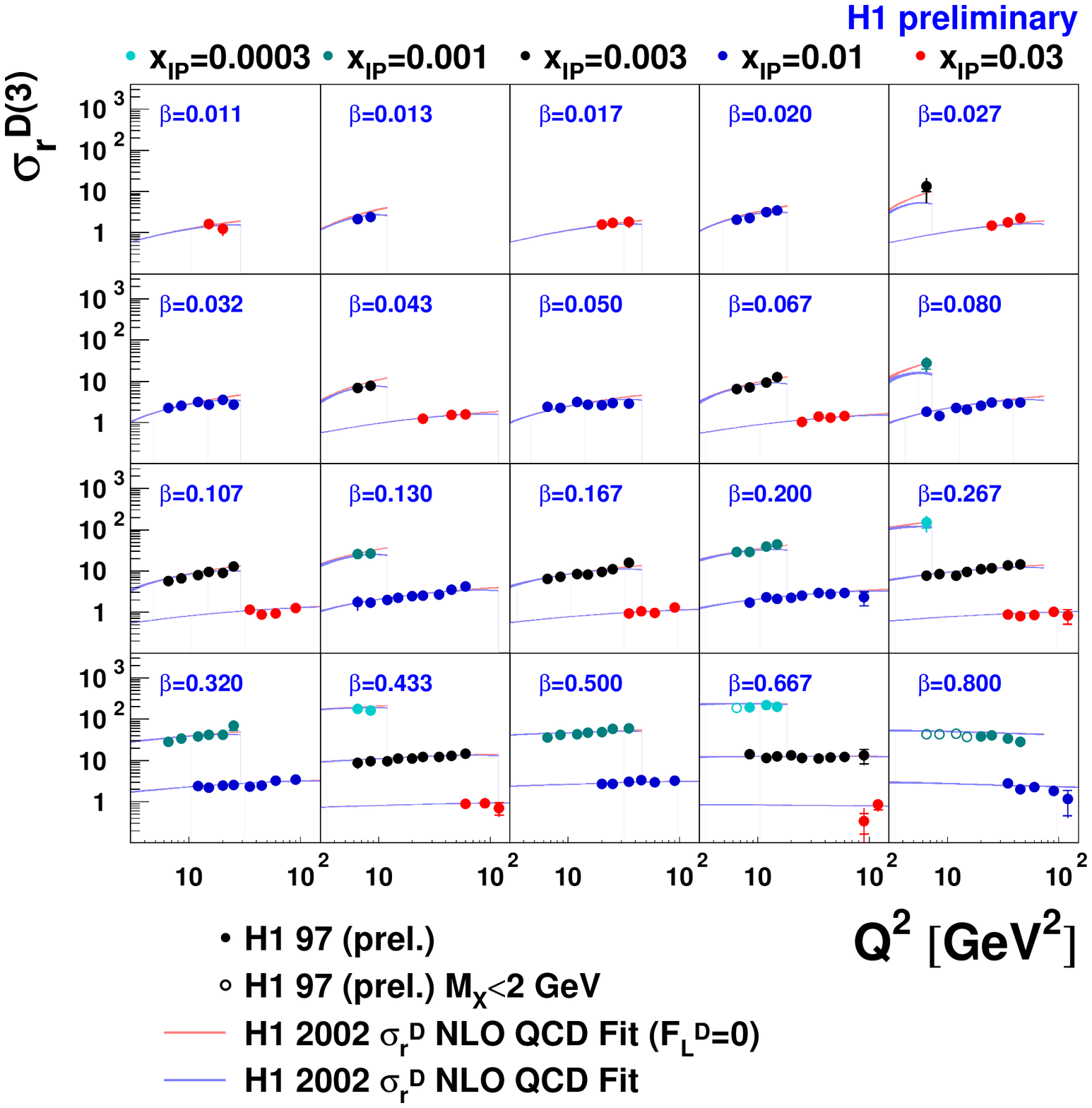,width=0.49\linewidth}
\caption{
The diffractive reduced cross section data as a function of $\beta$ (left) 
and $Q^2$ (right), compared with the result of the NLO QCD fit.}
\label{betaall}
\end{figure}

\begin{figure}[ht]
\centering
\epsfig{file=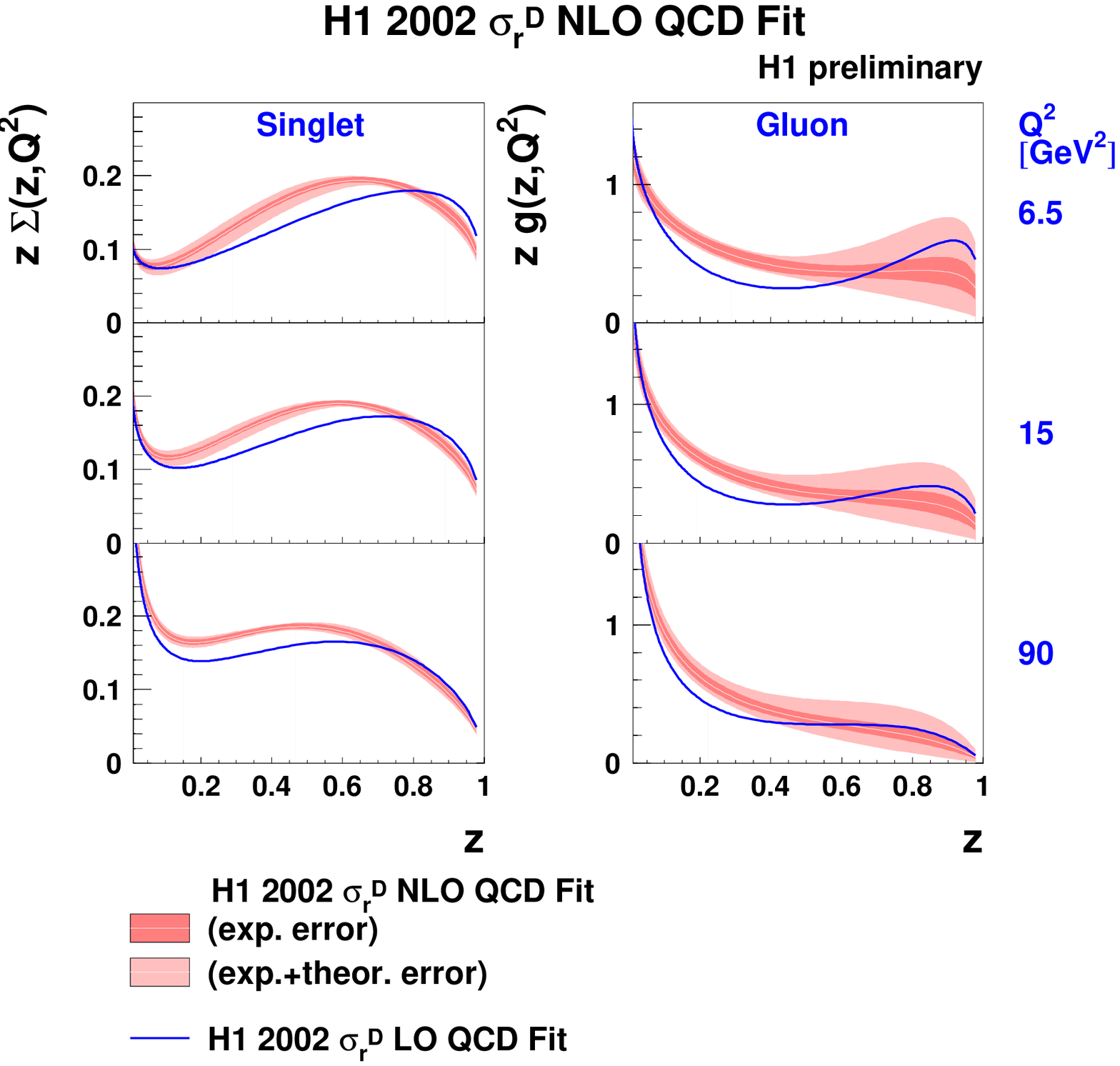,width=0.49\linewidth}
\epsfig{file=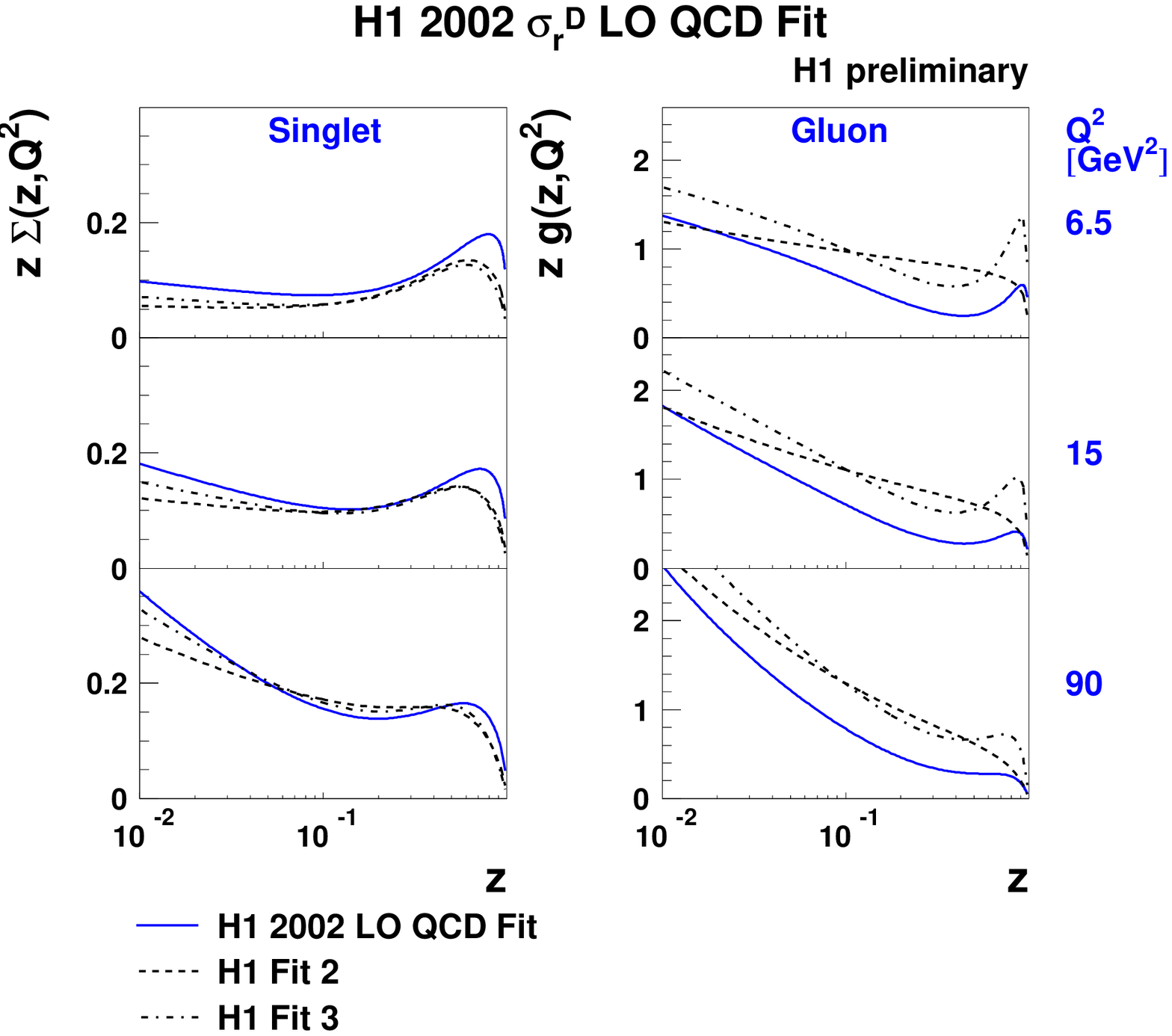,width=0.49\linewidth}
\caption{The dpdf's resulting from the NLO (left) and LO (right) QCD fits.}
\label{nlofit}
\end{figure}

\begin{figure}[ht]
\centering
\epsfig{bbllx=10,bblly=261,bburx=265,bbury=535,
file=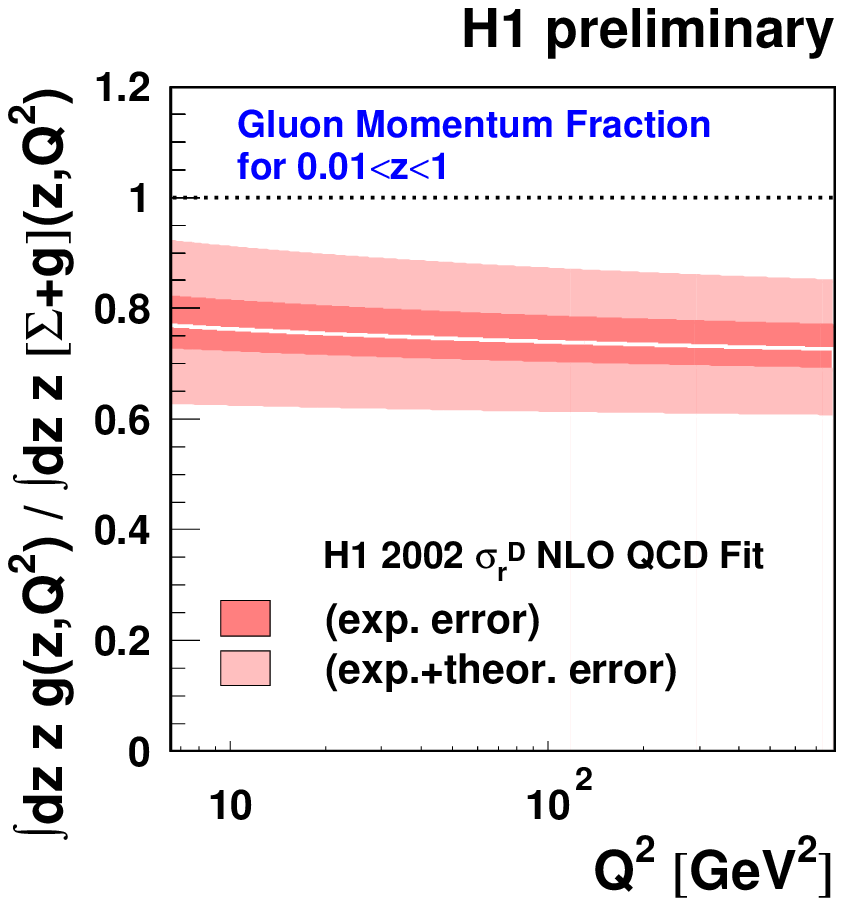,width=0.4\linewidth}
\epsfig{bbllx=0,bblly=355,bburx=290,bbury=720,
file=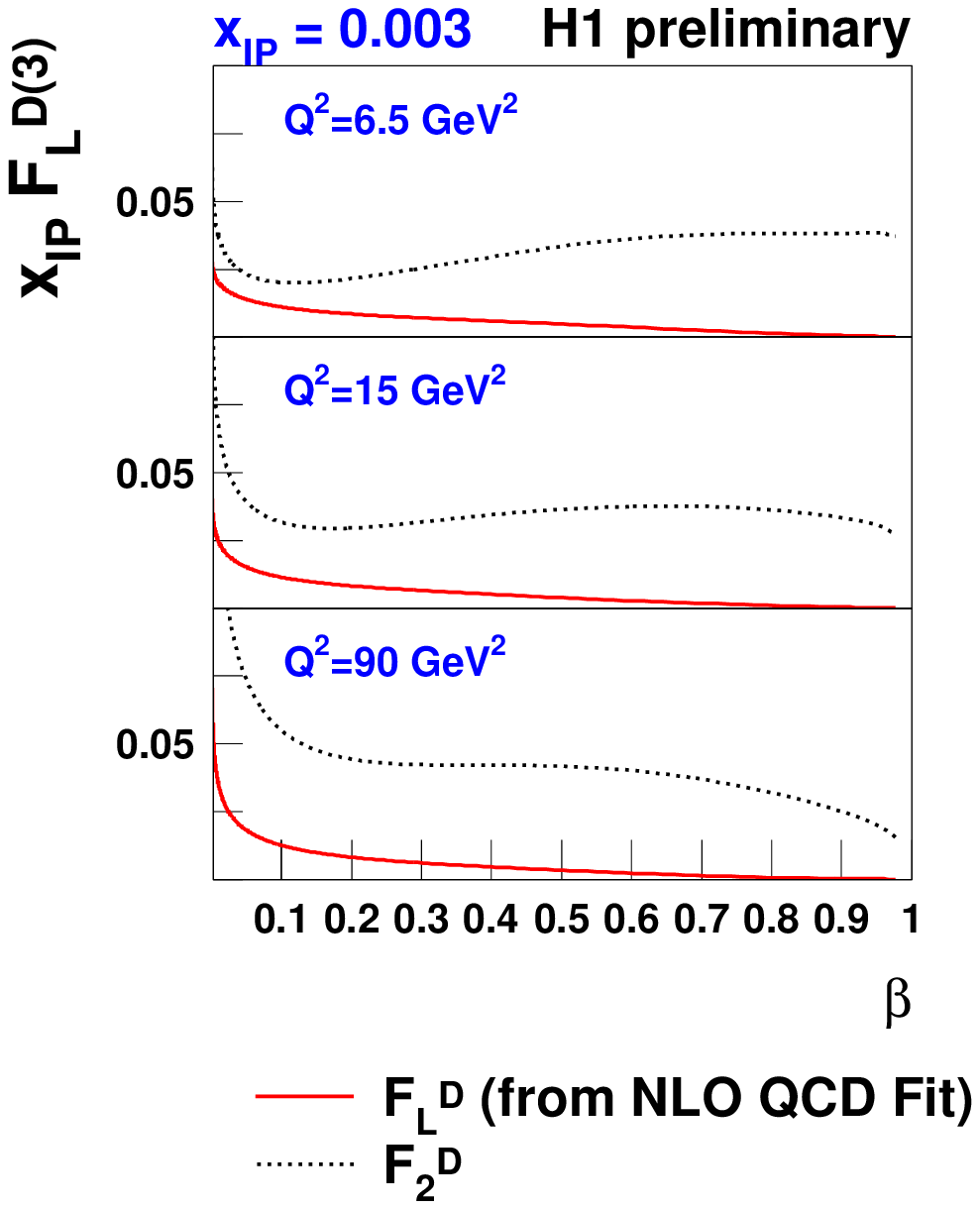,width=0.4\linewidth,height=0.425\linewidth}
\caption{Momentum fraction of the colour singlet exchange
carried by gluons (left) and the perturbative, leading twist 
part of $F_L^D$ (right) resulting from the NLO QCD fit.}
\label{gluonfl}
\end{figure}

\begin{figure}[ht]
\centering
\epsfig{file=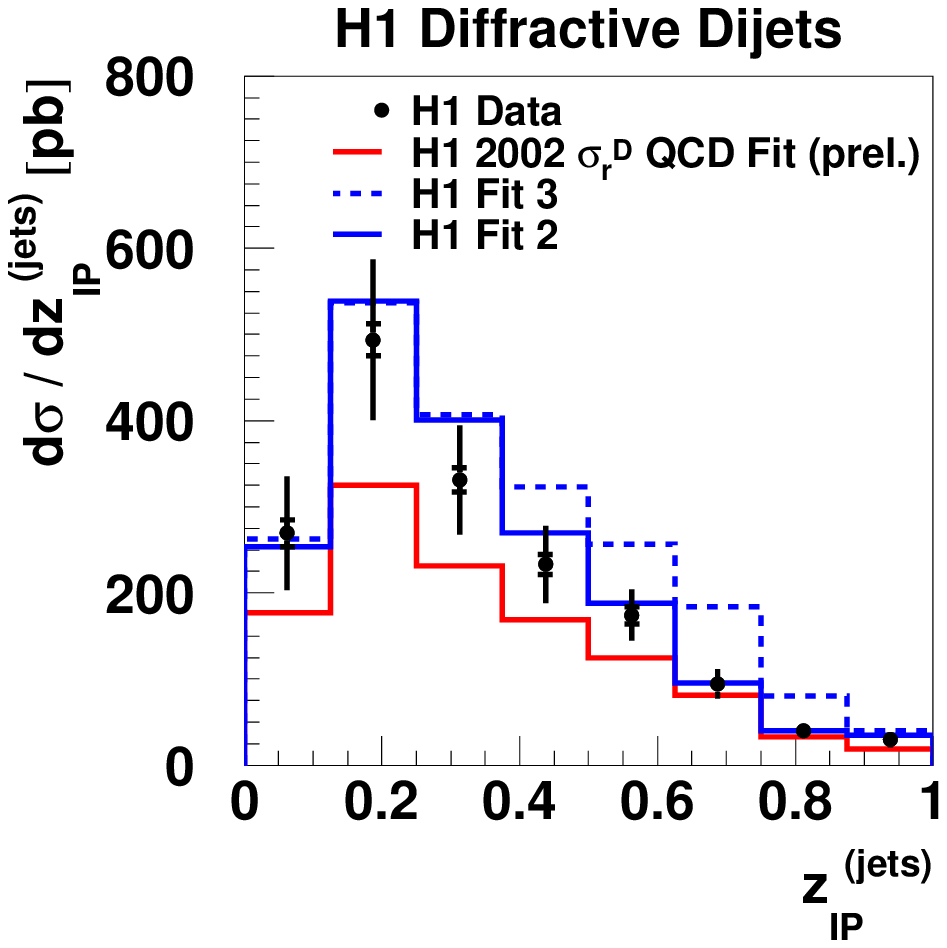,width=0.4\linewidth}
\epsfig{file=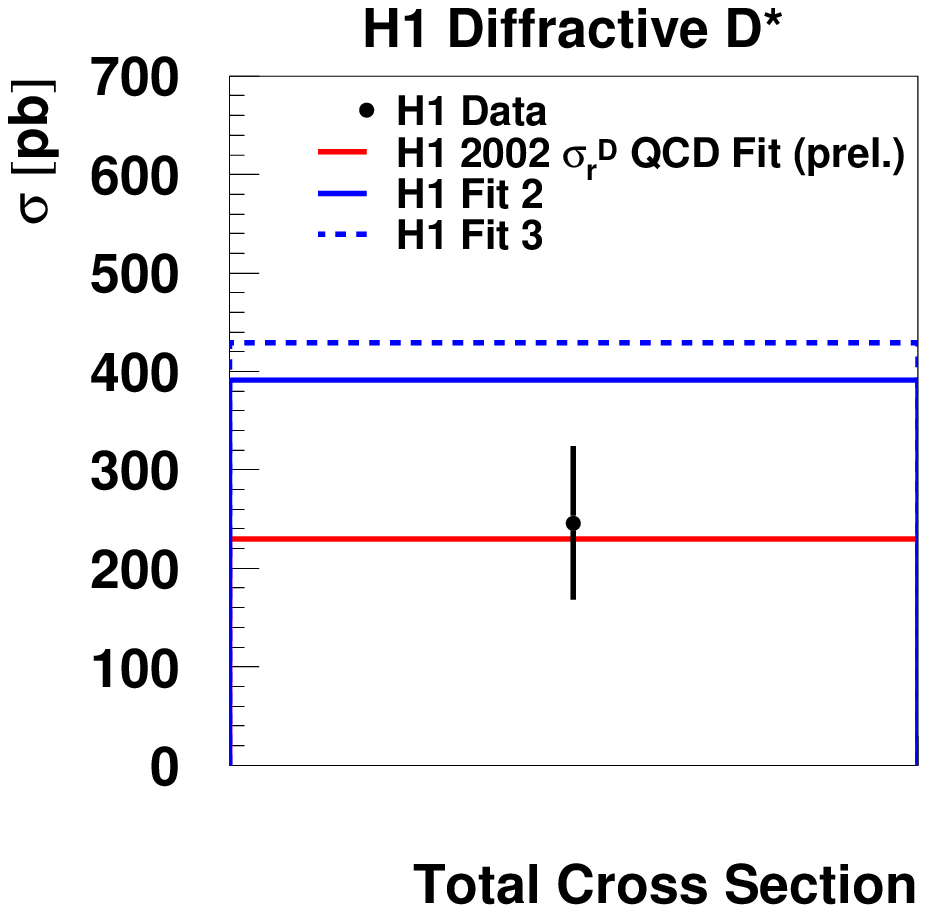,width=0.4\linewidth}
\caption{Comparison of H1 measurements of dijet
(left) and $D^*$ (right) cross sections in diffractive DIS with
predictions based on the new LO dpdf's.
}
\label{jetscharm}
\end{figure}

The result of the fit is compared with the data in Fig.~\ref{betaall}.
The diffractive parton distributions are presented in
Fig.~\ref{nlofit}, where the inner error bands correspond to the
experimental error and the outer error bands include in addition the
theory error.  The parton distributions extend to large fractional
momenta $z$ (or $\beta$). The gluon distribution is dominant. Fits at
leading order (LO) have been performed as well and are compared with
previous H1 fits \cite{h1f2d94} in Fig.~\ref{nlofit} (right). Taking
the uncertainties, especially of the old fits, into account, the
agreement is reasonable.

Integrating the parton distributions over the measured kinematic range
($0.01<z<1$), the momentum fraction of the colour singlet exchange
carried by gluons is obtained (Fig.~\ref{gluonfl} left). At $Q^2=10
\rm\ GeV^2$, it amounts to $75\pm15\%$. At NLO QCD, the perturbative
leading twist component of the longitudinal diffractive structure
function $F_L^D$ is predicted (Fig.~\ref{gluonfl} right). $F_L^D$ is
found to increase relative to $F_2^D$ towards low $Q^2$ and $\beta$.

\section{ Comparison with diffractive final state data}

QCD factorization in diffraction can be tested by using the dpdf's
extracted from $\sigma_r^D$ in DIS to predict the cross sections for
diffractive final states such as jet and charm production. Both
processes are strongly sensitive to the diffractive gluon
distribution.  Fig.~\ref{jetscharm} presents comparisons of H1 data
for diffractive dijet \cite{h1jets} and $D^*$ meson \cite{h1dstar}
production in DIS based on the LO dpdf's of the present as well as the
previous QCD fits\footnote{For more detailed comparisons, see
  \cite{thompson}.}.  The $D^*$ data are well described using the new
fit. Taking the uncertainties in the dpdf's and also the scale
uncertainties due to the LO comparison into account, the data are
consistent with QCD factorization.

\begin{wrapfigure}[16]{L}{0.5\linewidth}
\centering
\epsfig{clip=,file=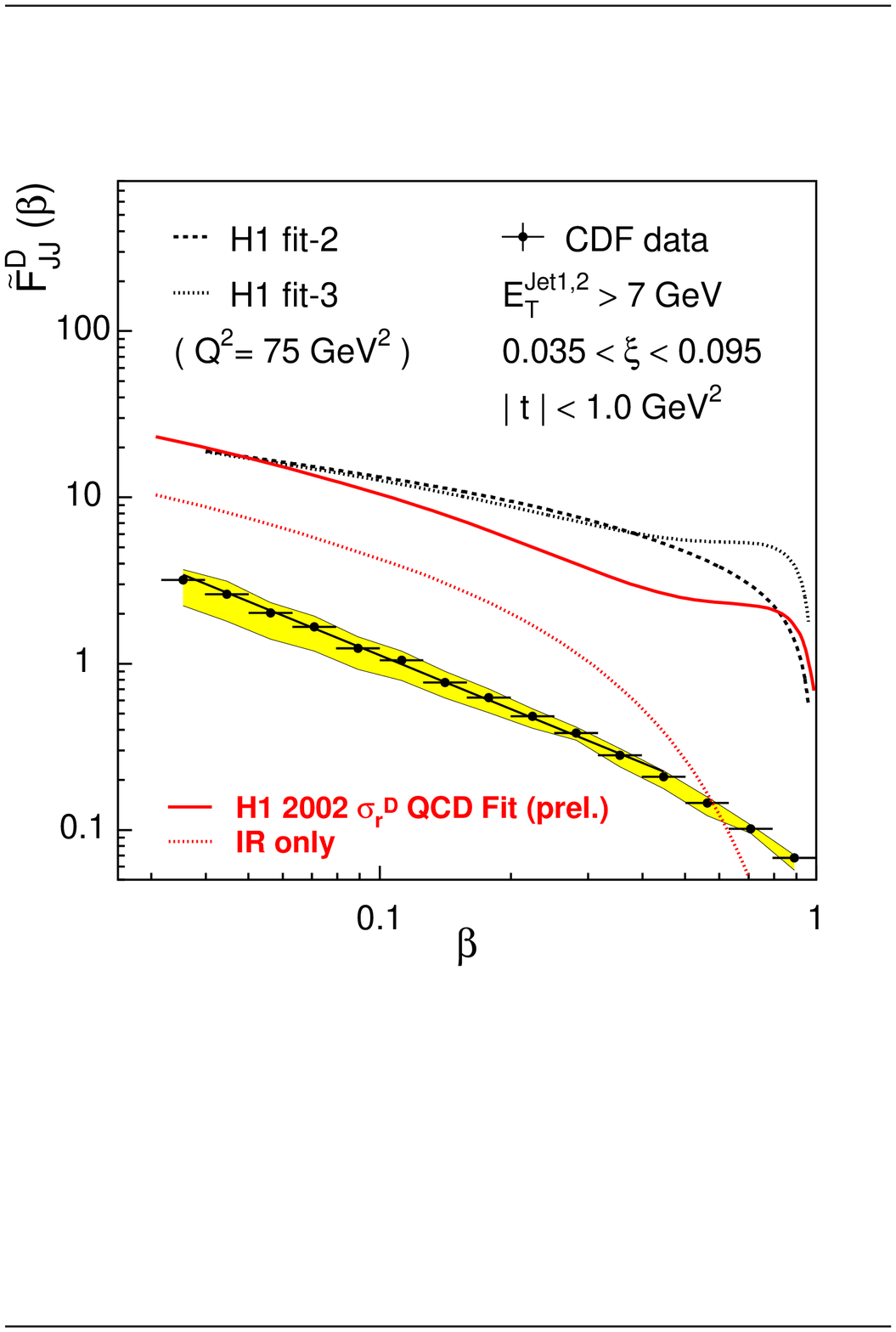,width=0.95\linewidth}
\caption{Comparison of CDF diffractive dijet data with a prediction
based on the new dpdf's and assuming factorization to hold.}
\label{cdf}
\end{wrapfigure}

Fig.~\ref{cdf} shows a comparison of CDF data for diffractive dijet
production at the TEVATRON with a prediction based on the dpdf's
obtained at HERA. The new QCD fit confirms the serious breakdown of
factorization in diffraction observed at hadron-hadron colliders,
often interpreted as being due to additional spectator interactions
which suppress rapidity gaps (absorptive corrections, ``rapidity gap
survival'').

\section{Conclusions}

Diffractive parton distributions were extracted from a NLO QCD
fit to recent H1 inclusive diffractive DIS data. For the first time in
diffraction, the uncertainties of these pdf's were evaluated.
Comparisons with diffractive final state data show, within the present
uncertainties, consistency with
QCD factorization in diffractive DIS and confirm the breakdown
of factorization at the TEVATRON.

\end{document}